\documentclass[prb,twocolumn,preprintnumbers,amsmath,amssymb]{revtex4-1}
\usepackage{graphicx}
\UseRawInputEncoding

\begin{document}
\title{  A  quantum  Kolmogorov-Arnold-Moser  theorem in the anisotropic Dicke model
         and its possible implications in the hybrid Sachdev-Ye-Kitaev models }
\author{  Yi-Xiang Yu$^{1}$, Jinwu Ye $^{2,3}$, Wuming Liu$^{4} $ and CunLin Zhang $^{5}$ }
\affiliation{
   $^{1}$  University of Science and Technology Beijing, Beijing 100083, China \\
   $^{2}$  Institute for Theoretical Sciences, Westlake University, Hangzhou, 310024, Zhejiang, China  \\
   $^{3}$  Department of Physics and Astronomy, Mississippi State  University, P. O. Box 5167, Mississippi State, MS, 39762   \\
   $^{4}$  Beijing National Laboratory for Condensed Matter Physics,
   Institute of Physics, Chinese Academy of Sciences, Beijing 100190, China  \\
   $^{5}$  Key Laboratory of Terahertz Optoelectronics, Ministry of Education and  Beijing Advanced innovation Center for Imaging Technology, Department of Physics, Capital Normal University, Beijing 100048, China  }
\date{\today }


\begin{abstract}
  The classical  Kolmogorov-Arnold-Moser (KAM) theorem provides the underlying mechanism for the stability
  of the solar system under some small chaotic perturbations.  Despite many previous efforts,
  any quantum version of the KAM theorem remains elusive.
  In this work, we provide a quantum KAM theorem in the context of the anisotropic Dicke model which is the most
  important quantum optics model. It describes a single mode of photons coupled to $ N $ qubits with both
  a rotating wave (RW) term and a counter-RW (CRW) term. As the ratio of the CRW over the RW term increases from zero to one,
  the systems evolves from quantum integrable to quantum chaotic.
  We establish a quantum KAM theorem to characterize such a evolution quantitatively
  by both large $ N $ expansion and Random Matrix Theory and find agreement from
  the two complementary approaches. Connections and differences between the Dicke models and Sachdev-Ye-Kitaev (SYK)
  or hybrid SYK models are examined.
  Possible Quantum KAM theorem in terms of other quantum chaos criterion such as quantum Lyapunov exponent is also discussed.
\end{abstract}
\maketitle

\section{Introduction}
    In classical chaos, the  Kolmogorov-Arnold-Moser (KAM) theorem \cite{KAM} describes how
    an integrable Hamiltonian $ H_0 $ respond to a chaotic perturbation $ \Delta H $,
    which makes the total Hamiltonian $ H=H_0+ \Delta H $ non-integrable. It states that if the two conditions are satisfied:
    (a) $  \Delta H $ is sufficiently small (b) the frequencies $ \omega_i $ of $ H_0 $ are in-commensurate, then the system remains quasi-integrable. The classical KAM theorem has played important roles in the stability
    of the solar system and many other classical chaotic systems.
    It remains an outstanding problem to find a quantum analogue of the KAM theorem for a quantum many-body system.
  Here, we will try to achieve such a goal in the context of the
  anisotropic (  $  J-U(1)/Z_2 $ ) Dicke  model \cite{walls,scully} Eq.\ref{u1z2u1} which is the most important model in quantum optics.

    There are previous works \cite{chaos,dicke,gold,comment,strongED,stupid}
    studying both quantum phase transitions (QPT) and quantum chaos in several extreme limits of the
    $  J-U(1)/Z_2 $ Dicke model Eq.\ref{u1z2u1}. It describes a single mode of photons coupled to $ N $ qubits with both
    a rotating wave (RW) $ g $  term and a counter-RW (CRW) $ g^{\prime} $ term at any ratio $ \beta=g/g^{\prime} $.
    For example, the authors in \cite{chaos} studied the  $ J-Z_2 $ Dicke model with $ \beta=1 $ at the thermodynamic limit $ J= \infty $ and  also its energy level statistics (ELS) \cite{WD1,WD2} at a finite $ J=N/2 $.
    In the $ J =\infty $ limit,
    as the atom-photon coupling strength increases above a critical value,
    it displays a QPT from the normal to the superradiant phase \cite{aue,subirbook}.
    However, the system  becomes non-integrable at any finite $ J $. By studying its ELS
    by exact diagonization (ED) at finite sizes $ N \geq 10 $ at a given parity sector,
    they found that in the normal phase, it is Possionian  $ P_{p}\left( s\right) =e^{-s} $,
    but in the superradiant phase, becomes Wigner-Dyson (WD) distribution  in the Gaussian orthogonal ensemble (GOE) class $P_{w}\left(s\right) =\frac{\pi }{2}se^{-\frac{\pi }{4}s^{2}}$ in the RMT classification \cite{WD1,WD2}.
    This fact suggests that the quantum Chaotic to Integrable transition (CIT)
    at a finite $ N $ may be associated to the QPT at  $ N=\infty $.
    On the other limit, the $ U(1) $ Dicke model \cite{dicke,gold} with $ \beta=0 $ is always integrable and still undergoes
    a QPT from the normal to the super-radiant phase at $ N=\infty $. This fact indicates that a QPT may not be related to a CIT.
    In \cite{gold,comment}, we evaluate  the whole energy spectrum of the $ J-U(1) $ Dicke model ( with $ J=N/2 $ ) with $ \beta=0 $
    by a $1/J $ expansion and find nearly perfect agreements with those found from the ED when $ N $ is even as small as $ N=2 $.


    It was well known that quantum dynamics are inherently encoded in any quantum many body systems,
    one effective way to characterize any possible quantum chaos in such an intrinsic quantum dynamics
    is through  Random matrix theory (RMT)\cite{WD1,WD2,ratiosta}.  We first propose a quantum version of KAM theorem
    to describe the quantum Chaotic to Integrable transition ( CIT ) in terms of the RMT. Then
  we derive the effective Hamiltonian Eq.\ref{u1z2u1} at any ratio $ 0 < \beta=g^{\prime}/g < 1 $  by the $ 1/N $ expansion.
  By using the effective Hamiltonian, we  investigate the analytic scaling form of the quantum KAM theorem at a finite $ N $ near
  its quantum integrable $ U(1) $ limit $ \beta=g^{\prime}/g \ll 1 $.
  By carefully identifying the chaotic source leading to the energy level correlations
  near the quantum integrable $ U(1) $ limit,
  we show that the quantum KAM theorem Eq.\ref{kam} holds
  in the strong coupling limit $ N > g/g_c \gg 1 $, so the system remains quasi-integrable.
  We stress the important roles played by the Berry phase  in its establishment.
  By using the RMT, we evaluate the energy level statistic (ELS) at $ N=10,20 $ at a given  parity sector by exact diagonizations (ED)
  at any $ 0 < \beta < 1 $.
  The ED data in the RMT fit well the analytic quantum KAM theorem Eq.\ref{kam} achieved by the $1/N $ expansion.
  We explore the intrinsic connections  between the QPT characterized by $ 1/N $ expansions
  and the CIT characterized by the RMT. Contrasts to the  Sachdev-Ye-Kitaev (SYK) and hybrid SYK models
  \cite{MBLSPT,randomM,period,KAMSYK,GangTian} are made.  Possible quantum KAM in terms of the Lyapunov exponent is discussed.

\section{  The $ 1/J $ expansion of the $ J-U(1)/Z_2 $ Dicke model in the super-radiant phase and QPT.}
  In the $ J-U(1)/Z_2 $ Dicke model \cite{walls,scully}, a single mode of photons couple to
  $  N $ two level atoms projected in the total angular momentum $ J=N/2 $  state \cite{gold,strongED}:
\begin{eqnarray}
  H_{J} &  = &  \omega_a a^{\dagger} a + \omega_b  J_{z}
  + \frac{g}{\sqrt{2J}} (  a^{\dagger} J_{-}+ a  J_{+}  )
      \nonumber   \\
  &  + &   \frac{g^{\prime}}{\sqrt{2J}} (  a^{\dagger} J_{+}+ a J_{-} )
\label{u1z2u1}
\end{eqnarray}
   where $ \omega_a, \omega_b $ are the energy of the cavity photons and the two atomic levels  respectively,
   $  g= \sqrt{N} \tilde{g}, g^{\prime}= \sqrt{N} \tilde{g}^{\prime} $ are the collective photon-atom rotating wave (RW) coupling
   and counter-rotating wave (CRW) term respectively .
   If $ \beta=g^{\prime}/g=0 $,  Eqn.\ref{u1z2u1} reduces to the $ U(1) $ Dicke model
   \cite{dicke} with the $ U(1) $ symmetry
   $ a \rightarrow  a  e^{ i \theta}, \sigma^{-} \rightarrow \sigma^{-} e^{ i \theta} $ leading to the conserved quantity
   $ P=  a^{\dagger} a + J_z $ where $ J^{z}= \frac{1}{2} \sum_{i} \sigma^{z}_{i} $.
   The CRW $ g^{\prime} $ term breaks the $ U(1) $ to the $ Z_2 $ symmetry
   $ a \rightarrow -a , \sigma^{-} \rightarrow -\sigma^{-} $ with the conserved parity operator  $ \Pi= e^{ i \pi ( a^{\dagger} a + J_z ) } $.
   If $ \beta=1  $, it becomes the $ Z_2 $ Dicke model \cite{chaos,strongED}.
   If $ \beta=\infty  $, it  can be mapped to the static version of Landau-Zener (LZ) model \cite{LZ0}.
   In this work, we fix the ratio to be $ 0< g^{\prime}/g = \beta < 1  $.
   The other case with  $ 1< \beta < \infty  $ need a different treatment and
   will be discussed  in a separate publication \cite{stupid}.

   Inside the super-radiant phase,  it is convenient to write
   both the photon and atom in the polar coordinates $ a= \sqrt{
   \lambda^{2}_{a} + \delta \rho_a } e^{ i \theta_a}, b= \sqrt{
   \lambda^{2}_{b} + \delta \rho_b } e^{ i \theta_b} $.
   We first minimize the ground state energy at the order $ J $ and
   found the saddle point values of $ \lambda_a $ and $ \lambda_b $:
\begin{equation}
 \lambda_a  =  \frac{ g + g^{\prime} }{ \omega_a } \sqrt{ \frac{j}{2} ( 1 - \mu^{2} )
  },~~~~ \lambda_b= \sqrt{ j(1-\mu) }
\label{meanz2}
\end{equation}
   where $  \mu = \omega_a \omega_b/(g + g^{\prime} )^{2} $. In the superradiant phase, $ \mu < 1 $, so that
   $ g+g^{\prime} > g^{t}_c= \sqrt{\omega_a \omega_b} $.
   In the normal phase $ g+g^{\prime} < g^{t}_c $,
   one gets back to $  \lambda_a=\lambda_b = 0 $. At a fixed $ \beta $, the QPT happens at $ g_c= \frac{ \sqrt{\omega_a \omega_b} }{1 + \beta } $.


   Well inside the superradiant phase, $ \lambda^{2}_a \sim
   \lambda^{2}_b \sim J $, it is  convenient to introduce the $ \pm $ modes:
   $ \theta_{\pm}= (\theta_a \pm \theta_b)/2, \delta \rho_{\pm}= \delta \rho_a \pm \delta \rho_b,
   \lambda^{2}_{\pm}= \lambda^{2}_a \pm  \lambda^{2}_b $.
   The Berry phase in the $ + $ sector \cite{gold} can be defined as
\begin{equation}
 \lambda^{2}_{+}= P + \alpha
\label{BP}
\end{equation}
   where $ P=1,2,\cdots $ is the closest integer  to the $ \lambda^{2}_{+} $,
   so $ -1/2 < \alpha < 1/2 $.

   After shifting $ \theta_{\pm} \rightarrow \theta_{\pm} + \pi/2 $,  we reach the Hamiltonian to the order of $ 1/J $:
\begin{eqnarray}
 {\cal H}[ \delta \rho_{\pm}, \theta_{\pm} ] & = &
 \frac{D}{2} (\delta \rho_{+} - \alpha )^{2} + D_{-} [\delta \rho_{-} + \gamma  ( \delta \rho_{+}-\alpha )]^{2}
               \nonumber    \\
   & + &  4 \omega_{a} \lambda^{2}_a [ \frac{ 1 }{ 1+ \beta } \sin^{2} \theta_{-}
     +  \frac{ \beta }{ 1+ \beta } \sin^{2} \theta_{+} ]
\label{pmu1z2h}
\end{eqnarray}
   where $ D=  \frac{ 2 \omega_a (g+ g^{\prime} )^{2} }{  E^{2}_{H} N }  $ is the phase diffusion constant in the $ + $ sector,
   $  D_{-}= E^{2}_{H}/16 \lambda^{2}_{a} \omega_a $ with $  E^{2}_{H}= ( \omega_a+\omega_b)^{2} + 4 ( g+ g^{\prime} ) ^2 \lambda^{2}_{a}/N $.
   The $ \gamma= \frac{ \omega^{2}_{a} }{ E^{2}_{H} } ( 1- \frac{ ( g + g^{\prime} )^{4}}{ \omega^{4}_{a} } ) $ is the coupling between the $ + $ and $ - $ sector. Due to the large gap in the $ \theta_{-} $ sector when $ 0< \beta < 1 $,
   it is justified to drop the Berry phase in the $ - $ sector.

  It is instructive to re-write Eq.\ref{pmu1z2h} as:
\begin{equation}
 {\cal H}[ \delta \rho_{\pm}, \theta_{\pm} ]  = H_{ U(1) }+  4 \omega_{a} \lambda^{2}_a \frac{ \beta }{ 1+ \beta }\sin^{2} \theta_{+}
\label{u1z2}
\end{equation}
     where the $ H_{U(1) } $ is the Hamiltonian of the $ J-U(1) $ model:
\begin{eqnarray}
    H_{ U(1) }= \frac{ D }{2} (\delta \rho_{+} - \alpha )^{2} ~~~~~~~~~~~~~~~~~~    \nonumber  \\
     + D_{-} [\delta \rho_{-} + \gamma  ( \delta \rho_{+}-\alpha )]^{2}
    +   4 \omega_{a} \lambda^{2}_a \frac{ 1 }{ 1+ \beta } \sin^{2} \theta_{-}
\label{u1h}
\end{eqnarray}
   which conserves $ \delta \rho_{+} $. Its eigen-energies and eigen-states are listed in \cite{gold}
   and also reviewed in the appendix A.
   Eq.\ref{u1z2} naturally separates the chaotic perturbation $ H^{\prime}_c  $ from those integrable ones.
   Near the integrable $ U(1) $ limit $ \beta \ll 1 $, the second term  $ H^{\prime}_c  $ in Eq.\ref{u1z2} can be treated as
   the small chaotic perturbation $ [ H^{\prime}_c, H_{ U(1) } ] \neq 0 $, it  violates the conservation of $ \delta \rho_{+} $,
   but still keeps the parity $ \Pi=e^{ i \pi ( P+\delta \rho_{+}) } $.
   Despite Eq.\ref{u1h} explicitly contains $ \beta $ dependencies, they still keep the integrability, so
   do not change the ELS. This observation  will be analyzed further in the following section.



\section{ A quantum KAM theorem: general statement.  }
    Inspired by the classical KAM theorem,
    we expect a quantum analogue of the KAM theorem exists near an integrable quantum many-body
    system whose eigen-energies are in-commensurate.
    In Eq.\ref{u1h}, it is the frustration due to the Berry phase $ \alpha $
    which make its eigen-energies in-commensurate except at $ \alpha=0, \pm 1/2 $
    which have zero measures anyway.
 We state the general form of a quantum KAM theorem from the RMT point of view:
 {\sl Near the integrable limit of an incommensurate quantum many body system,
 when the energy level repulsion caused by a small
 chaotic perturbation is less than the average many-body energy spacing of the integrable system,
 the system remains quasi-integrable, so its ELS remains to be Possionian. }
 This is justified, because all the energy levels in the integrable side are un-correlated.
 Specifically, taking two nearest  neighbour (NN) bulk energy states with the NN energy level spacing $ s_0= E^0_2-E^0_1 $,
 one can write down the $ 2 \times 2 $ quantum chaotic matrix within the two NN energy levels subspace as
 $ \Delta_{ij}= \langle i | H^{\prime}_c  | j \rangle, i,j=1,2 $.  By a bulk energy level $ E_B $, we mean
 $ \lim_{N\rightarrow \infty} \frac{ E_B- E_0 }{N} \neq 0 $ where $ E_0 $ is the ground state energy.
 Then the perturbed NN energy level spacing is:
\begin{equation}
 S= \sqrt{ s^2 + |  \Delta_{12} |^2 }
\label{NNshift}
\end{equation}
where $ s=s_0 + \Delta_{22}-\Delta_{11} $ is the diagonal energy shift due to the
chaotic perturbation, the $ \Delta_{12} $ is the off-diagonal one.
It is important to observe that if setting $ \Delta_{12} =0 $ in Eq.\ref{NNshift}, then the ELS still stays Possionian.
This is because the diagonal shift does not change the ELS, only the off-diagonal does.
Indeed, near any integral limit $ {\cal H}_0 $, if one adds a perturbation $ {\cal H}_1 $  which commutes with the
integral Hamiltonian $ [ {\cal H}_0, {\cal H}_1 ]=0 $,  $ {\cal H}_1 $  is a conserved quantity, the system remains integrable.
However, $ {\cal H}_1 $ still induces a diagonal energy shift, but not the off-diagonal one.
Obviously, it is the off-diagonal one which introduces the level repulsion between the two NN levels, which, in turns, leads to
the change of ELS to WD.  So we conclude that  when $  | \Delta_{12} | < s $ in Eq.\ref{NNshift},
the ELS stays Possionian, the quantum KAM applies.
In the following, instead of giving a rigorous mathematical proof of this quantum KAM theorem,
we derive its analytic finite size scaling form by $ 1/N $ expansion in Eq.\ref{u1z2}  and compare with our ED at various avaliable values of $ N $.

\begin{figure}
\includegraphics[width=7cm]{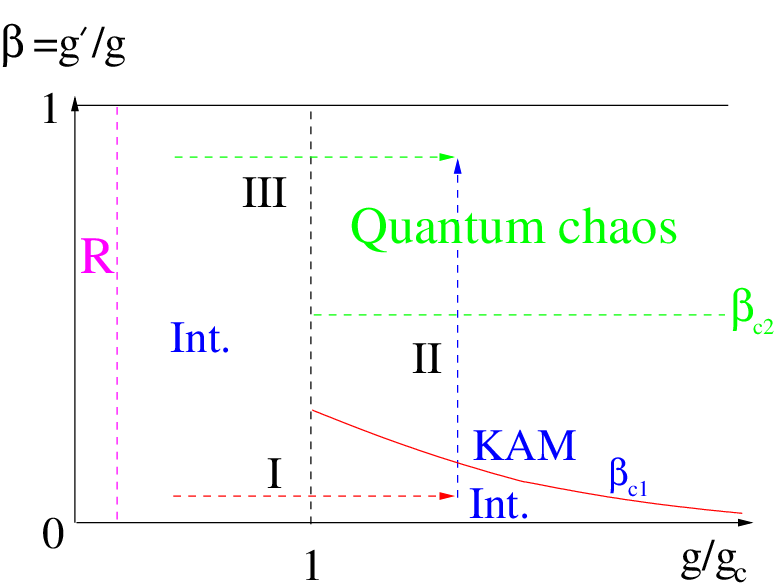}
\caption{ (Color online) QPT versus CIT in the large $ N $ limit. The $ U(1) $ limit $ \beta=0 $ is integrable,
only QPT. In the $ Z_2 $ limit $ \beta=1 $, the QPT is accompanied by the CIT. The quantum KAM scaling $ \beta_{c1} \sim N^{-2} $
Eq.\ref{kam}
achieved by $1/N $ expansion match qualitatively with that extracted from the ED in RMT in Fig.2.
 $ \beta_{c2} $ sets the quantum chaotic regime ( or dual KAM ) which remains finite as $ N \rightarrow \infty $.
 R one the left means the regular regime\cite{regular}.
 Path I: QPT tuned by the coupling $ g/g_c $, No CIT.
 Path II: CIT tuned by $ \beta $, No QPT.
 Path III: The QPT is accompanied by the CIT. }
\label{kam12}
\end{figure}

\section{ The quantum KAM theorem: the $ 1/N $ expansion on the Dicke model. }
 Let's take two NN states as $  |B1 \rangle= | l \rangle_m | m \rangle $
 and  $  |B2 \rangle= | l \rangle_{m+2} | m+2 \rangle $
 where  $ l \sim N/2 $ and $ m \sim 1 $ with a typical Berry phase $ -1/2 < \alpha < 1/2 $.
 Then $ \lim_{N\rightarrow \infty} \frac{ E_B- E_0 }{N} = \hbar \omega_0/2 $.
 One can immediately evaluate the diagonal matrix element $ \langle B1 | H^{\prime} |B1 \rangle
 = \langle B2 | H^{\prime} |B2 \rangle=2 \omega_{a} \lambda^{2}_a \frac{ \beta }{ 1+ \beta } $.
 So the chaotic perturbation does not change the diagonal energy level spacing
\begin{equation}
 s=E_0(l,m+2)-E_0(l,m)=2D(m-\alpha+2)
\label{dia}
\end{equation}
  which is independent of the Landau level index $ l \sim N/2 $.
  This fact simplifies the computation considerably.

  One can also compute the splitting ( NN energy level repulsion ) in terms of the coherent state:
\begin{equation}
 \Delta_{12}=\langle B1 | H^{\prime} |B2 \rangle
  =\omega_{a} \lambda^{2}_a \frac{ \beta }{ 1+ \beta }   \ _{m}\langle l| l \rangle_{m+2}
\label{off}
\end{equation}
  where one can evaluate the matrix element
\begin{eqnarray}
    f(l,l)&= & \ _{m}\langle l| l \rangle_{m+2}=\langle l| D(iG) | l \rangle
    \nonumber  \\
     & = & e^{-G^{2}/2} l ! \sum^{ l }_{r=0} \frac{ (-1)^{l-r}  G^{2l-2r} }
    { [(l- r ) !]^2 r ! }
\label{flk}
\end{eqnarray}
   where $ G=g_{m+2}-g_m= \frac{ \sqrt{2} \gamma }{ \beta_o } $.
   One can find $ f(0,0)= e^{-G^2/2}, f(1,1)=e^{-G^2/2} (1-G^2), \cdots $
   In fact, more straightforwardly, in terms of the wavefunction of a harmonic oscillator,
   $ f(l,l)= \int d \theta_{-} |\psi_{l} ( \theta_{-} )|^2 e^{ i 2 \gamma  \theta_{-} },l=1,2,3,\cdots $, one reach the same results as Eq.\ref{flk}.

   One can write down the general expressions of the three quantities
   $ \lambda^2_a $ in Eq.\ref{meanz2}, the diffusion constant $ D $ in Eq.\ref{pmu1z2h}
   and $ G^2/2 $ in Eq.\ref{flk} in terms of $ g/g_c >1 $ inside the super-radiant phase.
   They simplify dramatically in the strong coupling limit $ N > g/g_c \gg 1 $
\begin{eqnarray}
   \lambda^2_a &=&  (\frac{g}{g_c})^{2} N, \nonumber \\
   D &= & 2\omega (\frac{g}{g_c})^{-2}\frac{1}{N} \sim 2\omega/\lambda^2_a, \nonumber \\
   G^2/2 &= &\frac{\sqrt{1+\beta}}{2} \frac{1}{N}
\label{strongG}
\end{eqnarray}
 In the large $ N $ limit, the polynomial in $ f(l=N/2,l=N/2) $ multiplying  the exponential
 $ e^{-G^2/2} $ becomes $ (1-\frac{1}{2}+ \frac{1}{36}-\frac{1}{8\times36}+\cdots ) $.
 So we conclude $ f(N/2,N/2 ) \sim e^{-G^2/2}/2 $ as $ N \rightarrow \infty $.

 Applying the general criterion  to the two typical bulks states in Eq.\ref{dia} and Eq.\ref{off}
 leads to scaling form of the quantum KAM theorem:
 \begin{equation}
     \beta_{c1} \sim N^{-2}  e^{1/2N} (g/g_c)^{-4}
\label{kam}
 \end{equation}
   which approaches zero in both thermodynamic limit $ N \rightarrow \infty $ and the strong coupling limit $ N > g/g_c \gg 1 $.


  One may also propose the quantum KAM theorem from a dual point of view, namely from
  the chaotic $ Z_2 $ limit at $ \beta=1 $ inside the super-radiant phase in Fig.1 and Fig.2, namely,
  investigate how the chaotic behaviours change to integrable as the perturbation $ 1-\beta $ increases.
  There is no change of the symmetry as $ 1-\beta $ turns on,
  the stability of quantum chaos is much more robust than the KAM in the integrable side.
  If we define  $ \beta_{c2} $ as the dual form  of the KAM theorem in Fig.2, then $ \beta_{c2} $ remains finite
  as $ N \rightarrow \infty $.
  Inside the super-radiant phase $ g/g_c \geq 1 $, when  $ \beta_{c2} < \beta < 1 $, it remains chaotic.
  When $ \beta_{c1}  < \beta < \beta_{c2}  $, the ELS is in a crossover regime and satisfies neither GOE nor Possionian.
  When $ 0  < \beta < \beta_{c1}  $, it reaches the quantum KAM regime.

\begin{figure}
\includegraphics[width=9cm]{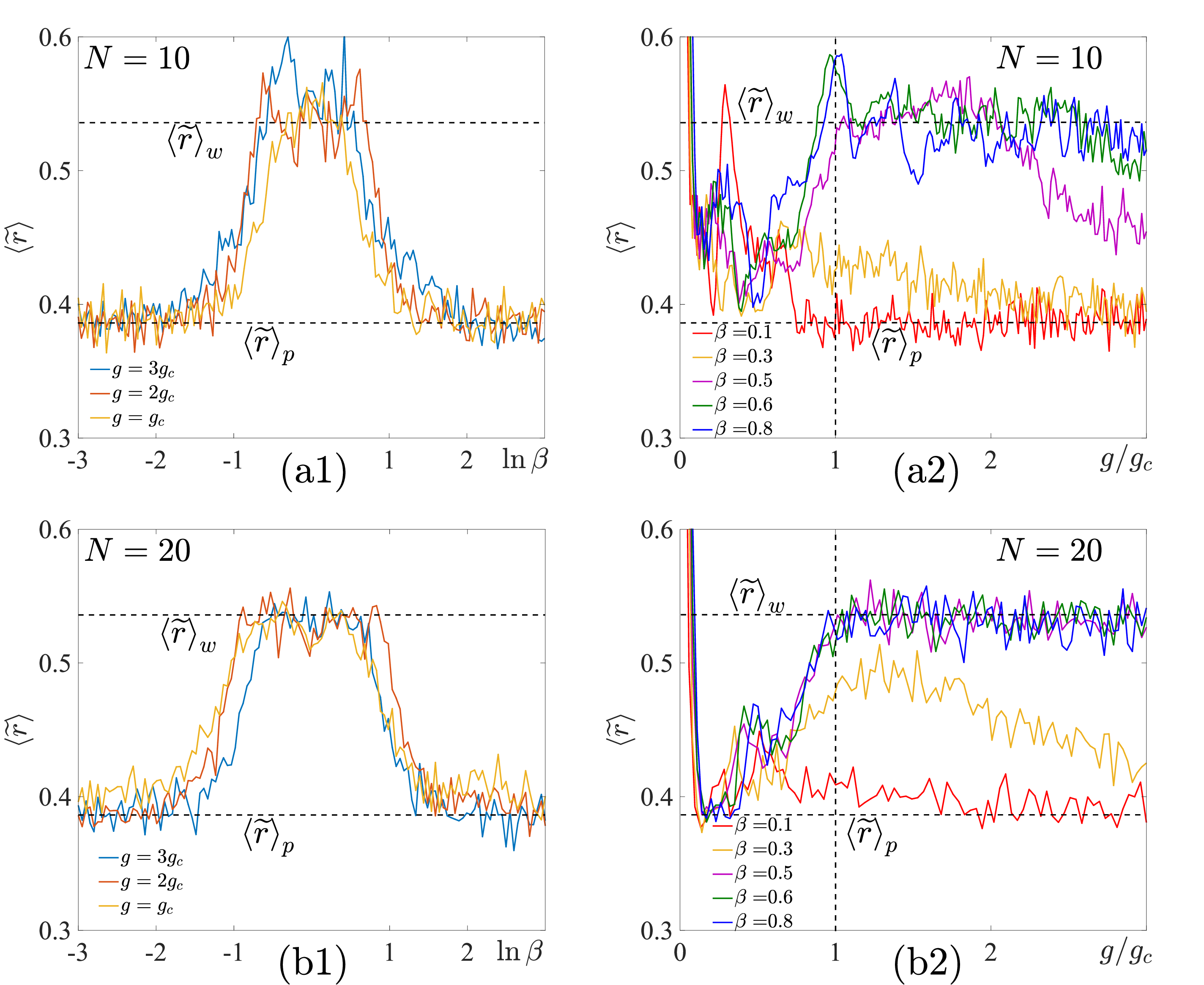}
\caption{ The mean values of the ratio $\widetilde{r}$ evaluated by ED at
(a) $N=10$ and (b) $ N=20 $ at the resonant case $\protect\omega_a=\protect\omega_b$.
The off-resonant case can also be similarly discussed.
The analytical mean values of $\widetilde{r} $ of Possion
$ \langle \widetilde{r} \rangle_p $ and GOE $ \langle \widetilde{r} \rangle_w $ are marked as references.
(a1) The CIT at a fixed $ g/g_c \geq 1 $ at three values $ g/g_c=3,2,1 $ and tuning $\protect\ln\beta$.
See path II in Fig.\ref{kam12}.
One can identify the KAM regime as $ \ln \beta_{c1} \sim \pm 1.2 $ which leads to $ \beta_{c1} \sim 0.3 $ near $ g/g_c > 1 $.
One may also identify the stability regime of the quantum chaos as $ \ln \beta_{c2} \sim \pm 0.5 $ which leads to
$ \beta_{c2} \sim 0.6 $.
When $ \beta_{c1} < \beta < \beta_{c2} $, there is no well defined ELS.
(a2) At a fixed $\beta $ versus $ g/g_c $ at 5 different values  $ \beta=0.1, 0.3, 0.5, 0.6, 0.8 $.
See path III and I in Fig.\ref{kam12}. One may also identify of $ \beta_{c1}\sim 0.3, \beta_{c2} \sim 0.6 $.
When $  \beta_{c2} < \beta < 1 $ ( path III in Fig.\ref{kam12} ), there is an accompanying                                                      CIT near the QPT at  $ g/g_c =1 $,
but when $ 0 < \beta < \beta_{c1} $ ( path I in Fig.\ref{kam12}), no such accompanying transition.
In both cases, when $  g/g_c < 0.15 $, $\widetilde{r}$ shoots up to $ 1 $,
the system just tends to be regular instead of being random \cite{regular}.
(b) The ED data at $ N=20 $ has less noises due to the larger finite size. Compared to $ N=10 $ in (a), one can see that
(b1) When $ g/g_c \geq 1 $, the quantum chaos regime expands, the integrable regime shrinks.
     One can identify $ \beta_{c2} \sim 0.5 $.
(b2) At the same set of $ \beta $ as (a2). The line with the same color ( $ \beta $ ) is elevated towards the
chaotic side, indicating $ \beta_{c1} $ drops as $ N $ increases. One can see $ \beta_{c1} \sim 0.1 $ near $ g/g_c > 1 $.
Our ED results also show $ \beta_{c1} $ decreases at a fixed $ N=20 $ and much larger $ g/g_c \sim 10-20 \gg 1 $ also in qualitative agreement with Eq.\ref{kam}. }
\label{CIT10}
\end{figure}

\section{ The energy level statistics in RMT and the CIT.}
  Now we look at RMT classification of  Eq.\ref{u1z2u1}. The Time reversal symmetry is nothing but the
  complex conjugate operator $ {\cal T}=K $ which acts as $ K a K^{-1}=a,  K a^{\dagger} K^{-1}=a^{\dagger}, K i K^{-1}=-i $.
  Obviously, $ K J_{a} K^{-1}= J_{a}, a=\pm, z $. It keeps the commutation relations $ [a,a^{\dagger} ]=1,
  [J_z, J_{\pm}]=\pm J_{\pm} $. So the many-body Hamiltonian Eq.\ref{u1z2u1} has the Time-reversal symmetry,
  also $ K^2=1 $, so it satisfies GOE.

  As shown in \cite{ratiosta,MBLSPT,randomM,period,KAMSYK} and reviewed in the appendix B, the most effective way to characterize an ELS is
  to study the distribution of the ratio of two NN energy level spacings $\left\langle \widetilde{r}\right\rangle $.
We  plot $\left\langle \widetilde{r}\right\rangle $ vs $ \ln \beta $
at a fixed $ g/g_c=3,2,1 $  and $\left\langle \widetilde{r}\right\rangle $ vs $ g/g_c $
at a fixed $ \beta=0.1, \cdots 0.8  $ in Fig.2a and Fig.2b respectively for two different sizes $ N=10,20 $.
The data $ \beta_{c1}\sim 0.3, 0.1 $ at the two different sizes $ N=10, 20$ match qualitatively the scaling in Eq.\ref{kam}.
A small discrepancy maybe attribute to the cutoff introduced in the ED ( See appendix C ).





\section{ Contrast anisotropic Dicke models with hybrid SYK models }




So far we focused on the $ U(1)/Z_2 $ anisotropic Dicke model which is the most important model in quantum optics.
The quantum chaos in the Dicke model is investigated here by both $1/N $ expansion and RMT.
Similar approaches have also been used to explore the quantum chaos in
the Sachdev-Ye-Kitaev (SYK) model which maybe dual to a quantum black hole  \cite{SY,Kit,Mald,GangTian,MBLSPT,randomM,period,KAMSYK}.
It is constructive to contrast the  (anisotropic) Dicke model to the ( hybrid ) SYK models.
   (a) Both have an infinite-range interaction, so are effectively $ 0+1 $ dimensional systems.
(b) Both show quantum chaos and quantum information scramblings which can be studied by $ 1/N $ expansion and Random matrix theory (RMT) respectively. But the mechanism leading to the quantum chaos is very much different.
The former is due to both interactions and disorders. The latter is due to  the atom-photon interactions
(c) The former is an  interacting fermionic model with quenched disorders, while the latter is a clean interacting bosonic one consisting of
    $ N $ qubits interacting with photons. So when performing the ED in RMT, the Hilbert space in the former is automatically finite $ \sim 2^N $,
    while the latter is infinite, so a finite cut-off need to be introduced ( see the appendix C ).
(d) The ground state of SYK is a conformably invariant gapless QSL which
   leads to the maximal Lyapunov exponent $ \lambda_L= 2\pi/\beta $ by $ 1/N $ expansion
   when $ 1 <\beta J < N $.
   In the super-radiant phase in the Dicke model, there is a symmetry breaking in the $ N \rightarrow \infty $ limit,
   but the symmetry breaking is restored at a finite $ N $ by the quantum tunneling process \cite{strongED}.
   So it has a finite gap $ \sim \omega $ at a given parity sector in the strong coupling $ g/g_c \gg 1 $ limit,
   the quantum Lyapunov exponent $ \lambda_L=0 $ in the low temperature range $ 1 < \beta \omega < N $.
   However, we expect $ \lambda_L=g [1-(1-\beta)+\cdots ] $ when $ T \gg \omega $ and reach the maximum  at the $ Z_2 $ limit $ \beta=1 $
   and vanish when $ \beta < \beta_{c1} $.
 (e) Various hybrid SYK model \cite{KAMSYK} also hold various CITs from the quantum integrable side of the $ q=2 $ SYK to
     the quantum chaotic side of the $ q=4 $ SYK tuned by the ratio of the couplings of the two sides.
     We expect the general statement on a KAM theorem in Sec.III still apply to the hybrid SYK models.
     Some preliminary results on the scaling forms of a quantum KAM in the hybrid SYK contexts are presented in \cite{KAMSYK}.
     However, due to the quenched disorders, constructing a rigours quantum KAM  to
     characterize the CIT in the hybrid SYK models maybe more challenging, but important to pursue.

\section{ Conclusions }

In this work, we take a $ 2 \times 2 $ matrix spanned by two typical NN bulk energy levels to find
the quantum KAM scaling Eq.\ref{kam}. This is justified, because all the energy levels in the integrable side are un-correlated.
In principle, one may also take all the bulk energy levels. However, as shown in \cite{WD1} and reviewed in the appendix B,
by considering a $ L \times L =2\times2$  matrices,
Wigner derived a simple approximate expression for the distribution function $P(s)$ of the NN spacing for GOE, GUE and GSE respectively.
Despite this so called Wigner surmise was achieved  only for $ 2 \times 2 $ matrices,  it is in very good
agreement with the exact large-$ L $ expressions. We expect the similar thing happen here: the exact $ L \rightarrow \infty $
calculation will only slightly modify the pre-factor in Eq.\ref{kam}. The qualitative agreement between our analytical KAM scaling Eq.\ref{kam}
with the ED data in the RMT supports this claim. It is important to extend the concepts and methods
developed here for interacting  bosonic systems to interacting fermionic systems with quenched disorders such as the hybrid SYK models.

It remains outstanding to construct a quantum KAM theorem through the Lyapunov exponent
or any other criterion characterizing the quantum chaos in the context of both Dicke and SYK models.
It is also interesting to establish some quantum-classical correspondence between the classical KAM
and the quantum KAM in the context of driven non-equilibrium Dicke \cite{drivenDicke} or driven non-equilibrium SYK models \cite{drivenSYK}.

{\bf Acknowledgements }

J Ye thank Fadi Sun for inspiring discussions on RMT and
the collaboration on the possible quantum analog of KAM theorem in the various hybrid SYK models \cite{KAMSYK}.
W.L were supported by the NKBRSFC under grants Nos. 2011CB921502, 2012CB821305, NSFC under grants Nos. 61227902, 61378017, 11311120053.
CLZ's work has been supported by National Keystone Basic Research Program (973 Program) under Grant No. 2007CB310408,
No. 2006CB302901 and by the Funding Project for Academic
Human Resources Development in Institutions of Higher Learning Under the Jurisdiction of Beijing Municipality.

\appendix

\section{ The eigen-energy and eigen-functions in the $ 1/N $  expansion. }
   Let us start with Eq.M5.
   Obviously $ [ H_{ U(1) }, \delta \rho_{+}  ]=0 $, the $ \delta \rho_{+} $ is a conserved quantity, so one can find the
   simultaneous eigenstates of $ H_{ U(1) } $ and $  \delta \rho_{+} $ which will be achieved in the following.


   As explained in Sec.M2, the $ \theta_{-} $ is very massive, after pinning $ \theta_{-} $ around $ \theta_{-} \sim 0 $,
one can approximate $ \sin^{2} \theta_{-} \sim  \theta^{2}_{-} $, therefore one may ignore the Berry phase in the $ \theta_{-} $ sector,
then the above equation can be simplified to:
\begin{eqnarray}
    H_{ U(1) } & = & \frac{D}{2} (\delta \rho_{+} - \alpha )^{2}
     + D_{-} [\delta \rho_{-} + \gamma  ( \delta \rho_{+}-\alpha )]^{2}   \nonumber   \\
     & + &  2 \omega_{a} \lambda^{2}_a \frac{ 2 }{ 1+ \beta }( \theta_{-} )^{2}
\label{u1hsimple}
\end{eqnarray}
     whose wavefunctions can be written as $ | m \rangle | l \rangle_{m} $ where
\begin{equation}
     \langle \theta_{+} | m \rangle = \frac{1}{ \sqrt{2 \pi}} e^{ i m \theta_{+} },
     \langle \theta_{-} | l \rangle_{m}= e^{ i \gamma( m-\alpha) \theta_{-} } \psi_{l} ( \theta_{-} )
\label{u1wave}
\end{equation}
    where the $ \psi_{l} ( \theta_{-} ) $ is just the $ l-th $ wavefunction of a harmonic oscillator. So the total wavefunction
    $ \psi_{l,m}( \theta_{+},\theta_{-} )= \langle \theta_{+},\theta_{-} | m \rangle | l \rangle_{m} $ is
\begin{equation}
      \psi_{l,m}( \theta_{+},\theta_{-} )=\frac{1}{ \sqrt{2 \pi}} e^{ i ( m \theta_{+} + \gamma( m-\alpha) \theta_{-} ) } \psi_{l} ( \theta_{-} )
\label{u1wavet}
\end{equation}
 where the wavefunction is only periodic in $ 0 < \theta_{+} < 2 \pi $, the $  - \infty < \theta_{-} < \infty $ is treated as a continuous variable. The corresponding eigen-energy is:
\begin{equation}
     E_{0}( l, m) = ( l+ 1/2 ) \hbar  \omega_{o} + \frac{D}{2} ( m-\alpha )^{2}
\label{energy0}
\end{equation}
    where the $ \omega_{o}= E_{H}/\sqrt{1+\beta} $ where
    $  E^{2}_{H}= ( \omega_a+\omega_b)^{2} + 4 ( g+ g^{\prime} ) ^2 \lambda^{2}_{a}/N $ is defined below Eq.M4.

  It is important to observe that Eq.\ref{u1hsimple} still contains $ \beta $ dependence.
  Only setting $ \beta =0 $ in Eq.\ref{u1hsimple} recovers the $ 1/N $ expansion of the original Eq.M1.
  As stressed below Eq.M6, despite Eq.\ref{u1hsimple} explicitly contains $ \beta $ dependencies, they still keep the integrability, so
  do not change the ELS. So it is a good starting point to look at the effects of a chaotic perturbation.

  Note that  the Landau level index $ l=0,1,...,N $ ( $ N+1 $ Landau levels ) denotes the high energy
  Higgs type of excitation, while the magnetic number $ m=-P+l,-P+l+1,.... $ ( no upper bounds ) denotes the low energy
  Goldstone types of excitations. The total parity is $  \Pi= (-1)^{P+m} $ at the sector $ P $ where $ P \ge l+1 $ has no upper bound either.
  At a given Landau level $ l $ and a given sector $ P $, there are $ |m|= P-l \ge 1 $ crossings at the
  Berry phase $ \alpha=0 $ in Eq.\ref{energy0}.
  The $ m=0 $ is always the ground state energy at a given $ l $ and a given $ P \ge l+1 $.

    In the large $ J=N/2 $ limit $ \omega_o \gg  D \sim 1/j $,
    so the first term can be considered as the inter-Landau levels, while the second term
    can be considered as the intra-Landau levels.

     It is instructive to look at the harmonic oscillator in the $ - $ sector from a algebraic point of view.
    Defining  $ \beta_o=\sqrt{ \frac{ \mu \omega_{o}}{\hbar}} $ where $ \mu=\frac{1}{2 D_{-}} $
    is the mass of the harmonic oscillator. The annihilation operator
     $ a_{-}= \frac{1}{\sqrt{2}}( \beta_o \theta_{-} + i \frac{ \delta \rho_{-} }{ \beta_o \hbar} ) $.
    Then after making a momentum shift $ \delta \rho_{-} \rightarrow
    \delta \rho_{-} - \gamma  ( \delta \rho_{+}-\alpha ) $, the annihilation operator
    $ a_{-} \rightarrow a_{-,m}= a_{-} + i \frac{ \gamma ( m-\alpha ) }{ \sqrt{2} \beta_o } $,
    then the harmonic oscillator's Hamiltonian is $ H_{-}= ( a^{\dagger}_{-,m}a_{-,m} + 1/2 ) \hbar \omega_{o} $.
    Its eigenstate $  a^{\dagger}_{-,m}a_{-,m}| l \rangle_{m}= l | l \rangle_{m} $ is
\begin{equation}
      | l \rangle_{m}= D^{\dagger}( i g_{m} ) |l \rangle=D( - i g_{m} ) |l \rangle
\label{coh}
\end{equation}
    where $ g_{m}=\frac{ \gamma ( m-\alpha ) }{ \sqrt{2} \beta_o } $ and $ |l \rangle $ is just the $ l $-th
    the harmonic oscillator eigenstate.
    Particularly, the ground state $ | 0 \rangle_{m}= D( - i g_{m} ) |0 \rangle $  is a coherent state.

    One can show that $  D( - i g_{m} )= e^{ i  \gamma ( m-\alpha ) \theta_{-} } $, then
    $ | l \rangle_{m}= e^{ i \gamma( m-\alpha) \theta_{-} }|l \rangle $.
    So $  \langle \theta_{-} | l \rangle_{m}= e^{ i \gamma( m-\alpha) \theta_{-} }  \langle \theta_{-}|l \rangle
    = e^{ i \gamma( m-\alpha) \theta_{-} } \psi_{l} ( \theta_{-} ) $, so we recover Eqn.\ref{u1wave} from the coherent state.

\section{ A brief Review on Random matrix theory and the Energy level statistic (ELS) of
Nearest Neighbour (NN) energy level spacings.}

For the 3 Wigner-Dyson classes: A(GUE), AI(GOE), AII(GSE),
no mirror symmetry exists in the energy levels.
Let $\{E_n\}$ be an ordered set of energy levels, the joint probability distribution for all the eigen-values can be described by
\begin{equation}
P(\{E_i\})\propto\prod_{i<j}|E_i-E_j|^\beta
\prod_n e^{-E_n^2}
\label{npoint}
\end{equation}
where $\beta$ is the Wigner-Dyson index characterizing the strength of level repulsion.
For class A(GUE), AI(GOE), AII(GSE), $\beta=2,1,4$ respectively.

One can denote $s_n=E_{n+1}-E_{n}$ as the NN spacings.
By considering a $2\times2$ matrices in Eq.\ref{npoint},
Wigner \cite{WD1} derived a simple approximate expression for the distribution function $P(s)$ of the NN spacing,
\begin{align}
    P_{w,\beta}(s)=a_\beta s^\beta e^{-b_\beta s^2}
\label{PW}
\end{align}
where $\beta=1,2,4$ is the Dyson index for GOE, GUE and GSE respectively.
Despite this so called Wigner surmise was achieved only for $ 2 \times 2 $ matrices,  it is in very good
agreement with the exact large-N expressions.

It is also known that independent random energy levels would yield a Poisson distribution
\begin{align}
    P_{p}(s)=e^{-s}
\label{PP}
\end{align}
However, in order to compare different results from different systems,
the energy levels will need an unfolding procedure,
which is not convenient when large enough statistics is not available
To get rid of the dependence on the local density of states,
it is convenient to look at the distribution
of the ratio of two adjacent energy level spacings \cite{ratiosta,MBLSPT,randomM,period,KAMSYK}
$r_{n}=s_{n}/s_{n+1}$ which distributes around $1$.
This quantity has the advantage that it requires no unfolding
since ratios of consecutive level spacings are independent of the local density of states.

By considering  $3\times3$ matrices system,
the authors in \cite{ratiosta} obtained the Wigner-like surmises of
the ratio of consecutive level spacings distribution
\begin{equation}
    P_{p}(r)=  \frac{1}{(1+r)^2},
    \quad
    P_{w}(r)=  \frac{1}{Z_{\beta}} \frac{(r+r^2)^{\beta} }{(1+r+r^{2})^{1+3 \beta/2}}
\label{PWr}
\end{equation}
where $\beta=1,2,4$
and $Z_{\beta}=8/27, 4 \pi/81\sqrt{3},  4 \pi/729\sqrt{3}$
for GOE, GUE and GSE respectively.
The distribution $P_W(r) $ has the same level repulsion at small $r$ as $P_W(s)$,
namely, $P_W(r)\sim r^\beta$.
However, the large $r$ asymptotic behavior $P_W(r)\sim r^{-(2+\beta)}$ is
dramatically different than the fast exponential decay of $P_W(s)$.

One may also compute the distribution of the logarithmic ratio
\cite{ratiosta,MBLSPT}  $P(\ln r)=P\left( r\right) r$.
Because $P \left( \ln r\right)dr $ is symmetric under $ r \leftrightarrow 1/r $,
one may confine $0<r<1$ and double the probability density
$P\left( \tilde{r}\right) =2P\left(r\right) $. Therefore, the above two distributions have two different sets of expected values of
$\tilde{r} = \min \{r, 1/r \} $:
\begin{align}
    \langle\tilde{r}\rangle_{p}
	&=\int_{0}^{1}2 r P_{p}(r) dr=2\ln 2-1 \approx 0.38629 ,    \nonumber   \\
    \langle\tilde{r}\rangle_{w}
	&=\int_{0}^{1}2 r P_{w,\beta=1,2,4}(r) dr
\label{PWr2}
\end{align}%
which is $4-2\sqrt{3} \approx 0.53590$,
$2\sqrt{3}/\pi-1/2 \approx 0.60266$,
$32\sqrt{3}/(15\pi)-1/2 \approx 0.67617$
for GOE,GUE and GSE respectively. These Wigner-like surmises Eq.\ref{PWr},\ref{PWr2} were also shown to be very accurate when
compared to numerics and exact calculations in the exact large-N expressions.

In the main text, the CIT is from the GOE to the Possion, so only the two values
$ \langle\tilde{r}\rangle_{p}  \approx 0.38629 $ and $ \langle\tilde{r}\rangle_{GOE} \approx 0.53590 $
are used and plotted in Fig.M2.

\section{ The high energy cutoff in the  exact diagonizations (ED). }
We do the ED in Fig.M2 on the $ J-U(1)/Z_2 $ Dicke model Eq.M1 in the  Fock   basis where the complete basis is $ | n \rangle | j, m \rangle, n=0,1,2,.....\infty, j=N/2,
 m=-j,.....,j $  where the $ n $ is the number of photons and the $ | j, m \rangle
 $ is the Dicke states. In performing the ED, following Ref.5, one has to use a
 truncated basis $ n=0,1,......n_c $ in the photon sector  where the $ n_c \sim 500-2000 \gg N $ is the maximum photon number in the artificially truncated Hilbert space. The total number of states is $ n_c \times (2j+1)= n_c \times (N+1) $.
 This is also the size of the RMT $ L=  n_c \times (N+1) $.
 The average many body energy level spacing at a given parity sector
 is $ \frac{ n_{c} \omega_a }{ 2 n_c \times (N+1) } \sim   \frac{ \omega_a }{ 2(N+1) } $. This qualitative estimate is consistent with Eq.M8 achieved by the systematic $1/N $ expansion.
 As long as the energy levels in Fig.M2 are well below $ n_c \omega_a $, then the
 energy levels should be very close to the exact results without the truncation ( namely, sending $  n_c \rightarrow \infty $ ).
 However, the ED may not be precise anymore when $ g $ gets too close to the upper cutoff.







\begin{thebibliography}{99}

\bibitem{KAM}
A. N. Kolmogorov, "On the Conservation of Conditionally Periodic Motions under Small Perturbation of the Hamiltonian," Dokl. Akad. Nauk SSR 98 (1954);  V. I. Arnold, "Proof of a theorem of A. N. Kolmogorov on the preservation of conditionally periodic motions under a small perturbation of the Hamiltonian " Uspekhi Mat. Nauk 18 (1963); J. Moser, "On invariant curves of area-preserving mappings of an annulus," Nachr. Akad. Wiss. Gottingen Math.-Phys. Kl. II 1962 (1962), 1¨C20.

\bibitem{walls} D. F. Walls and G. J. Milburn, Quantum Optics,
Springer-Verlag, 1994.

\bibitem{scully} M. O. Scully and M. S. Zubairy, Quantum Optics,
Cambridge University press, 1997


\bibitem{chaos} C. Emary and T. Brandes, Phys. Rev. Lett. 90, 044101 (2003); Phys. Rev. E 67, 066203 (2003).
It was shown that only when $ N \geq 6 $, it make sense to talk about the ELS. For example, the $ N=1 $ case which is the Rabi model,
it makes no sense to study its ELS.

\bibitem{dicke} K. Hepp and E. H. Lieb, Ann. Phys. (NY) 76, 360 (1973); Y. K. Wang and F. T. Hioe, Phys. Rev. A 7, 831 (1973).
V. N. Popov and S. A. Fedotov, Sov. Phys. JETP 67, 535 (1988); V. N. Popov and V. S. Yarunin, Collective Effects in Quantum Statistics of Radiation and Matter (Kluwer Academic, Dordrecht, 1988).
V. Buzek, M. Orszag, and M. Rosko, Phys. Rev. Lett. 94, 163601 (2005).
Jinwu Ye  and  CunLin Zhang, Super-radiance, Photon condensation  and its phase diffusion,
        Phys. Rev. A 84, 023840 (2011).


\bibitem{gold} Yu Yi-Xiang, Jinwu Ye and W.M. Liu, Scientific Reports 3, 3476 (2013).

\bibitem{comment}  Yu Yi-Xiang, Jinwu Ye, W.M. Liu and CunLin Zhang, arXiv:1506.06382.


\bibitem{strongED} Yu Yi-Xiang, Jinwu Ye and CunLin Zhang, Parity oscillations and photon correlation functions in the $ Z_2/U(1) $ Dicke model at a finite number of atoms or qubits, Physical Review A 94.2 (2016), 023830.


\bibitem{stupid} There is also a recent numerical study on the quantum chaos in the $ J-U(1)/Z_2 $ Dicke model:
 Wouter Buijsman, Vladimir Gritsev, and Rudolf Sprik, Nonergodicity in the Anisotropic Dicke Model,
 Phys. Rev. Lett. 118, 080601 ¨C Published 23 February 2017.
 This purely numerical paper did not
 (1) perform any $ 1/N $ expansion at a finite $ N $,
 (2) touch the KAM theorem at a finite $ N $
 (3) explore the intrinsic relations between the CIT at a finite $ N $ and
 the normal to super-radiant QPT at $ N =\infty $.
 This paper also made a technical mistake to identify the $ 0 < \beta < 1 $ case with the $ 1 < \beta < \infty  $ case.
 As argued below Eq.\ref{u1z2u1}, the $ g=0 $ case in Eq.\ref{u1z2u1} can be mapped to the static version of Landau-Zener (LZ) model \cite{LZ0}.
 In this work, we fix the ratio to be $ 0< g^{\prime}/g = \beta < 1  $.
 The other case with  $ 1< \beta < \infty  $ need a completely different treatment and
 will be discussed  in a separate publication.


\bibitem{WD1}   E.P. Wigner,
	\emph{On the statistical distribution of the widths and spacings of nuclear resonance levels}
	Proc. Camb. Phil. Soc., No. {\bf 47}, 790 (1951).
\bibitem{WD2}   F. Dyson,
	\emph{Statistical Theory of the Energy Levels of Complex Systems. I}
	J. Math. Phys. (N.Y.) {\bf 3}, 140 (1962).

\bibitem{aue} A. Auerbach,
\textit{Interacting electrons and quantum magnetism},
(Springer Science \& Business Media, 1994).

\bibitem{subirbook}
S. Sachdev,
\textit{Quantum Phase transitions},
(2nd edition, Cambridge University Press, 2011).

\bibitem{LZ0} Alexander Altland, V. Gurarie, T. Kriecherbauer, and A. Polkovnikov,
Nonadiabaticity and large fluctuations in a many-particle Landau-Zener problem,
Phys. Rev. A 79, 042703 ¨C Published 2 April 2009.



\bibitem{ratiosta} Y. Y. Atas, et al. "Distribution of the ratio of
consecutive level spacings in random matrix ensembles." Physical review
letters 110.8 (2013): 084101.










\bibitem{MBLSPT} Y.-Z. You, A. W. W. Ludwig, and C. Xu, Sachdev-Ye-Kitaev Model and Thermalization on the
Boundary of Many-Body Localized Fermionic Symmetry Protected Topological States, Phys. Rev. B 95, 115150 (2017).


\bibitem{randomM} J. S. Cotler, G. Gur-Ari, M. Hanada, J. Polchinski, P. Saad, S. H. Shenker, D. Stanford, A. Streicher,
and M. Tezuka, Black Holes and Random Matrices," (2016), arXiv:1611.04650 [hep-th].


\bibitem{period}  Fadi Sun and Jinwu Ye,  Periodic Table of SYK and supersymmetric SYK, Phys. Rev. Lett. 124, 244101 (2020).

\bibitem{KAMSYK}   Fadi Sun, Yu Yi-Xiang, Jinwu Ye and WuMing Liu,
A new universal ratio in Random Matrix Theory and chaotic to integrable transition
in Type-I and Type-II hybrid Sachdev-Ye-Kitaev models, Phys. Rev. B 104, 235133 (2021).


\bibitem{regular} Note that in Fig.2a2,2b2, there is always a regular regime when $ g/g_c < 0.15 $.
This is because in Eq.1, as $ g \rightarrow 0 $,
at a fixed $ \beta=g^{\prime}/g $, $ g^{\prime} \rightarrow 0 $, then only the first two terms survive
$ H_{0}   =   \omega_a [ a^{\dagger} a + J_{z} ] $ which is clearly regular.





\bibitem{SY} S. Sachdev and J. Ye, Gapless spin liuid ground state in a random quantum Heisenberg magnet,"
Phys. Rev. Lett. 70, 3339 (1993), cond-mat/9212030.


\bibitem{Kit} A. Y. Kitaev, Talks at KITP, University of California, Santa Barbara," Entanglement in Strongly-
Correlated Quantum Matter (2015).



\bibitem{Mald} J. Maldacena and D. Stanford, Remarks on the Sachdev-Ye-Kitaev model," Phys. Rev. D 94, 106002
(2016), arXiv:1604.07818 [hep-th].

\bibitem{GangTian}
R. Feng, G. Tian, D. Wei,
\emph{Spectrum of SYK model}, arXiv:1801.10073. Peking Math J. 2 (2019), No. 1, 41-70.
\emph{Spectrum of SYK model II: Central limit theorem}, arXiv:1806.05714;
\emph{Spectrum of SYK model III: Large deviations and concentration of measures}, arXiv:1806.04701.
Random Matrices Theory Appl. 9 (2020), No. 2, 2050001 24pp. 60B20 (60F10)

\bibitem{drivenDicke} K. Baumann, R. Mottl, F. Brennecke, and T. Esslinger,
Exploring Symmetry Breaking at the Dicke Quantum,
Phase Transition, Phys. Rev. Lett. 107, 140402 (2011).
Despite the claim made by these authors, the Dicke model realized here is not the equilibrim Dicke one studied here.
It may be so in the rotating frame with the pumping laser, but it is a driven non-equilibrium Dicke model in the lab frame.

\bibitem{drivenSYK}  Clemens Kuhlenkamp, Michael Knap,
Periodically Driven Sachdev-Ye-Kitaev Models, Phys. Rev. Lett. 124, 106401 (2020).



























\end{thebibliography}
\end{document}